\documentclass[preprint,authoryear,12pt]{elsarticle}

 \usepackage{graphicx}

\usepackage{amssymb}

\usepackage[
a4paper=true,%
breaklinks=true,%
colorlinks=true,%
pdfauthor={A. Marino et al.},%
pdftitle={Template for manuscripts in Advances in Space Research}%
]{hyperref}

\journal{Advances in Space Research}

\begin{document}

\begin{frontmatter}



\title{Galaxy evolution in groups. \\USGC U268 and USGC U376 in the Leo cloud}
 


\author[1]{A. Marino}
\author[2]{L. Bianchi}
\author[3]{P. Mazzei}
\author[3]{R. Rampazzo}
\author[1]{G. Galletta}

\address[1]{Dipartimento di Fisica ed Astronomia ``Galileo Galilei'', Vicolo dell'Osservatorio, 3, 35122, Padova, Italy}
 
\cortext[]{Corresponding author: A. Marino}
\fntext[footnote2]{Email address: antonietta.marino@oapd.inaf.it}

 \address[2]{Dept. of Physics and Astronomy, Johns Hopkins University, 3400 North Charles Street, Baltimore, MD 21218, USA}
 
\address[3]{INAF-Osservatorio Astronomico di Padova, Vicolo dell'Osservatorio 5, 35122 Padova, Italy}

\begin{abstract}
With the aim of investigating galaxy evolution in nearby galaxy groups, 
we analysed the spectral energy distribution  of 24 galaxies, members of  two groups in the Leo cloud,
USGC U268 and USGC U376.  We estimated the ages and stellar masses of the galaxies 
by fitting their total apparent magnitudes from far-ultraviolet to near-infrared with population synthesis
models. 
The comparison of the results for a subsample of galaxies with smooth particle hydrodynamic (SPH) simulations
with chemo-photometric implementation,
shows that in most  cases the estimated stellar masses obtained with the two different approaches 
are in good agreement. 
The kinematical and dynamical analysis indicates that USGC U268 is in a pre-virial collapse phase
while USGC U376 is likely in a more evolved phase towards virialization.
 
\end{abstract}


\end{frontmatter}

\parindent=0.5 cm

\section{Introduction}

The physical processes marking  galaxy evolution 
leave their imprinting on the spectral energy distribution (SED).   
SEDs are consequently widely used for deriving the
basic properties of galaxies, e.g. the  mass and
the average age of the stellar component, assuming a 
star formation history (SFH)
\citep[e.g.][]{Searle73, Larson78, Salim07}.

This powerful diagnostic, applied to samples of galaxies in clusters, groups, and in the general field, 
can reveal   environmental effects 
on their properties. In this context, 
\citet{Marino10}  combined  ultraviolet data from the Galaxy Evolution Explorer
 \citep[GALEX,][]{Morrissey07}  and  optical data from the Sloan
Digital Sky Survey (SDSS)  to 
study the galaxies  of the ``Local Group analog"  LGG 225 in the Leo II cloud.  
Fitting the observed SEDs from far-UV (FUV)  to near-IR (NIR) with stellar
population models, the total stellar mass of the group  
was estimated between  5 and 35 $\times$ 10$^{10}$
M$_{\odot}$ and the ages of the stellar populations  from a few to $\approx$ 7 Gyr.

\begin{table*}
	\caption{Main characteristics of the galaxy sample$^a$.}	 
	 \tiny
	\begin{tabular}{lcccllccl}
	\hline\hline
	   Group  & RA & Dec.&  Morph.&   E(B-V)$^b$& Incl. &  Mean Hel.& B$_T$ \\ 
	    
	   Galaxy & (J2000) & (J2000) & type   &  &   &    velocity  & \\
		   & (h:m:s) & (d:m:s)   & &   [mag]  &[deg]  &   [km/s] & (AB mag)\\
				  
	\hline
	U268  & & & & & & & &    \\
	\hline
 MRK 0408   &  09 48 04.8 &+32 52 58 & S?      &0.017 & 40.5  &1470$\pm$40$^c$ & 14.84$\pm$0.48\\  
 NGC 3003   &  09 48 35.7 &+33 25 17 & Sbc     &0.013 & 90.0  &1498$\pm$36$^c$ & 12.09$\pm$0.10 \\  
 NGC 3011   &  09 49 41.2 &+32 13 16 & S0      &0.016 & 38.6  &1548$\pm$36$^c$ & 14.38$\pm$0.44\\  
 UGC 05287  &  09 51 28.1 &+32 56 35 & Sc      &0.014 & 41.2  &1482$\pm$42$^c$ & 14.68$\pm$0.42 \\  
 UGC 05326  &  09 55 24.5 &+33 15 47 & IB      &0.016 & 16.9  &1415$\pm$45$^c$ & 14.46$\pm$0.40 \\  
 IC 2524    &  09 57 33.0 &+33 37 11 & S0-a    &0.013 & 63.1  &1487$\pm$10$^c$ & 14.84$\pm$0.49 \\  
 NGC 3067   &  09 58 21.3 &+32 22 11 & SABa    &0.015 & 81.8  &1491$\pm$36$^c$ & 12.57$\pm$0.12 \\  
 UGC 05393  &  10 01 42.1 &+33 08 12 & SBd     &0.013 & 68.8  &1448$\pm$25$^c$ & 14.79$\pm$0.38 \\  
 UGC 05446  &  10 06 30.9 &+32 56 49 & Sc      &0.014 & 80.5  &1383$\pm$42$^c$ & 15.22$\pm$0.40 \\  
 NGC 3118   &  10 07 11.6 &+33 01 40 & Sbc    &0.018 & 90.0  &1315$\pm$37$^c$ & 14.18$\pm$0.35 \\  
 PGC2016633 	  &09 48 02.68 &+32 54 01.7	   & ?    &0.016   &   53.1 &  1552$\pm$47 &  17.09$\pm$0.50 \\
PGC2042146 	 & 09 49 03.1 &+33 59 28.95	   &?	 &0.010 &    44.5   &  1494$\pm$2     &  17.69$\pm$0.50 \\
SDSSJ094911.28+342634.2 &  09 49 11.28 &+34 26 34.2& ?    &0.012 &	 &	  1489$\pm$2 &  	 \\
SDSSJ094935.09+342616.3 & 09 49 35.09 &+34 26 16.3 & ?    & 0.010&	      &  1488$\pm$2 &	\\
PGC082546 	 & 09 50 20.9 &+33 35 02	   & Sc  &0.015 & 46.3 &    1586$\pm$29    &  16.89$\pm$0.36 \\
NGC3021 	 & 09 50 57.1 &+33 33 13	   & Sbc &0.014 & 55.7 &    1540$\pm$3     &  12.38$\pm$ 0.27 \\
UGC05282   &09 51 10.4 & +33 07 53		   & Sm  &0.014 &    67.6 &    1557$\pm$10    & 	\\
PGC2025214  &09 53 45.2 & +33 09 52.0  	   &	? &0.013 &    57.2 &    1520$\pm$35    &   17.92$\pm$ 0.50 \\
SDSSJ095430.02+320342.0  & 09 54 30.02 &+32 03 42.0	&  ?   & 	&	  &    1421$\pm$2     &        \\
SDSSJ100309.92+323622.5  & 10 03 09.92 &+32 36 22.5	&   ?  &0.015&	      &        1562$\pm$96    &        \\
PGC029522   &       10 08 58.0 & +32 00 38		   & Sc  &  0.018  &   57.7  &1468$\pm$34    &   16.75$\pm$ 0.44 \\

\hline \hline	
 											 	      		       
	U376  &  & & & & & & &   \\
\hline
NGC 3592      &11 14 27.5 &+17 15 34&  Sc   &  0.015 &  79.4 &1298$\pm$20$^c$  &  14.31$\pm$0.30 \\
NGC 3599      &11 15 27.0 &+18 06 37&  S0   &  0.021 &  28.3 &812$\pm$31$^c$  &  12.72$\pm$0.08 \\
NGC 3605      &11 16 46.6 &+18 01 01&  E    &  0.021 &  90.0 &695$\pm$39$^c$  &  13.00$\pm$0.28 \\
UGC 06296     &11 16 51.0 &+17 47 55$^b$&  I&  0.019 &  90.0 &909$\pm$26$^c$  &  14.02$\pm$0.41 \\
NGC 3607      &11 16 54.7 &+18 03 06&  E-SO &  0.021 &  34.9 &955$\pm$41$^c$  &  10.77$\pm$0.17 \\
NGC 3608      &11 16 59.0 &+18 08 55&  E    &  0.021 &  47.0 &1199$\pm$42$^c$ &  11.41$\pm$0.26 \\
CGCG 096-024  &11 17 58.0 &+17 26 29&	?    &  0.020 &  28.9 &807$\pm$66$^c$  &  14.94$\pm$0.29 \\
UGC 06320     &11 18 17.5 &+18 50 48&	 ?   &  0.023 &  20.3 &1125$\pm$5$^c$  &  13.67$\pm$0.41 \\
UGC 06324     &11 18 22.1 &+18 44 18&  S0   &  0.022 &  90.0 &1068$\pm$38$^c$ &  14.61$\pm$0.32 \\
NGC 3626     &11 20 03.9 &+18 21 24&  S0-a &  0.020 &  56.1 &1570$\pm$37$^c$ &  11.64$\pm$0.23 \\
NGC 3655     &11 22 54.7 &+16 35 24&  Sc   &  0.025 &  47.1 &1490$\pm$39$^c$ &  12.16$\pm$0.07 \\
NGC 3659     &11 23 45.3 &+17 49 05&  SBd  &  0.019 &  68.8 &1282$\pm$39$^c$ &  12.76$\pm$0.48 \\
NGC 3681     &11 26 29.8 &+16 51 48&  Sbc  &  0.026 &  15.2 &1236$\pm$53$^c$ &  12.26$\pm$0.16 \\
NGC 3684     &11 27 11.2 &+17 01 48&  Sbc  &  0.026 &  50.8 &1131$\pm$39$^c$ &  12.15$\pm$0.11 \\
NGC 3686     &11 27 44.1 &+17 13 26&  SBbc &  0.024 &  42.0 &1096$\pm$36$^c$ &  11.84$\pm$0.07 \\
NGC 3691     &11 28 09.4 &+16 55 14&  SBb  &  0.026 &  47.4 &1081$\pm$39$^c$ &  12.48$\pm$0.17 \\
PGC 034537  &  11 18 06.0  &+18 47 53       & Sc      &0.023 & 23.3  & 1140$\pm$7 & 16.19$\pm$0.38 \\
PGC 086629  &  11 18  21.4  &+17 41 51       & I      &0.021 & 0.0   & 1055$\pm$5 &	   \\
UGC 06341   &  11 20 00.7  &+18 15 38       & Sd      &0.019 & 90.0  & 1611$\pm$42& 15.96$\pm$0.39 \\
PGC1534499  &  11 21 25.41 &+17 30 35.16    & Sm      &0.024 & 65.5  & 979$\pm$57 & 16.30$\pm$0.50 \\
PGC 086673  &  11 22 59.3  &+17 28 27       & I       &0.021 & 0.0   & 1383$\pm$4 &	 \\
PGC 035087  &  11 25 01.8  &+17 05 09       & Sm      &0.025 &       & 1208$\pm$5 &	 \\
PGC035096   &  11 25 10.8   & +16 53 04     & I       &0.026&  51.0  & 1021$\pm$3 &  15.96$\pm$1.17\\
PGC 035426  &  11 29 54.4  &+16 25 46       &Im$^b$   &0.027 &       & 1067$\pm$5 &	\\

 \hline
\end{tabular}
 										   
$^a$data from \tt HYPERLEDA http://leda.univ-lyon1.fr \citep{Paturel03}. 					   
$^b$ Taken from NED. 												   
$^c$ Taken from \citet{Ramella02}.

 \label{tab1}
\end{table*}

We are extending this analysis by studying groups  
characterized by different galaxy populations in the same Leo cloud.   \citet{Marino13} 
revisited the group membership and  investigated the substructures  of 
two groups, USGC U268 and  USGC U376 (U268 and U376 hereafter). 

In this work we derive the age of the stellar population (for an assumed SFH), the stellar mass, 
and the additional foreground extinction of galaxies in these two groups
 by fitting their SED with  a grid of models
computed with GRASIL code \citep{Silva98}, adopting SFH for  
 elliptical, spiral and irregular galaxies  as defined in \citet[][their Table 5]{Marino10}. 
Together with the characterization of the  galaxy  and group properties,
the objective of the present study is to 
compare results from the analysis with GRASIL models  
with  those obtained from SPH simulations with chemo-photometric 
implementation, matching not only the  integrated-SED but also the galaxy's kinematical 
and morphological structure. This comparison validates the integrated-SED analysis approach, 
which can be applied to larger samples. 
   
The paper is organized as follows.  In Section 2 we present the observed
SEDs of Ellipticals, Spirals and Irregulars of U268 and U376. In Section 3
we use synthetic galaxy populations models to interpret the SEDs, estimate ages and masses and  compare 
SED-fitting results with   
SPH simulations. Kinematical and dynamical analysis of the groups is presented in section 4. 
Summary and conclusions are given in Section 5.  

 \begin{figure*}
 \begin{center}
 \includegraphics[width=10cm]{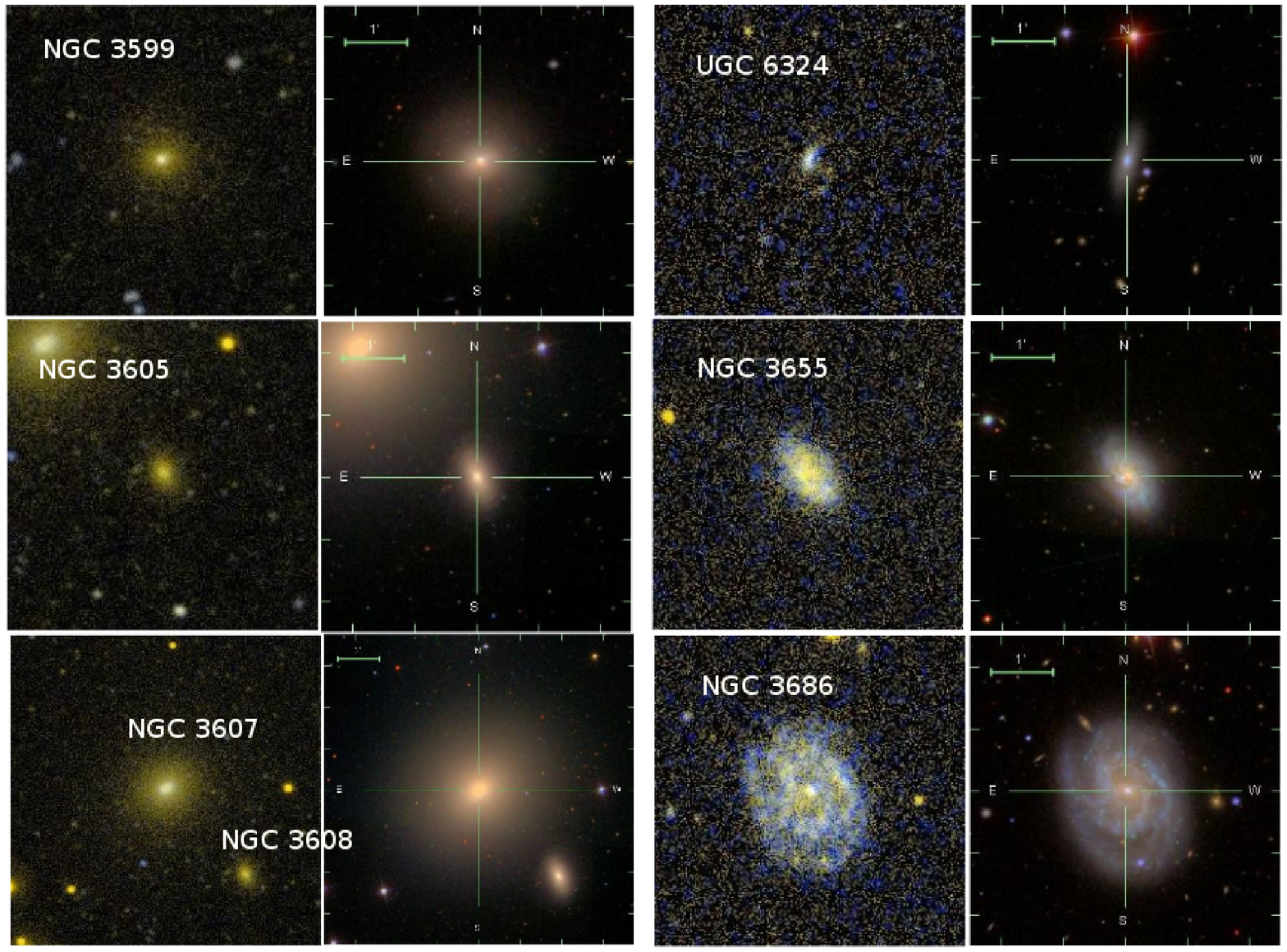}
        \end{center}
      \caption{Color composite UV (FUV blue, NUV yellow, columns 1 and 3), and optical (SDSS, g blue, r green, i red, columns 2 and 4)      images of 6 members of U376 discussed in Section 4.
}
       \label{f1}
   \end{figure*}

\begin{figure*}
 \begin{center}
 \includegraphics[width=12cm,angle=-90]{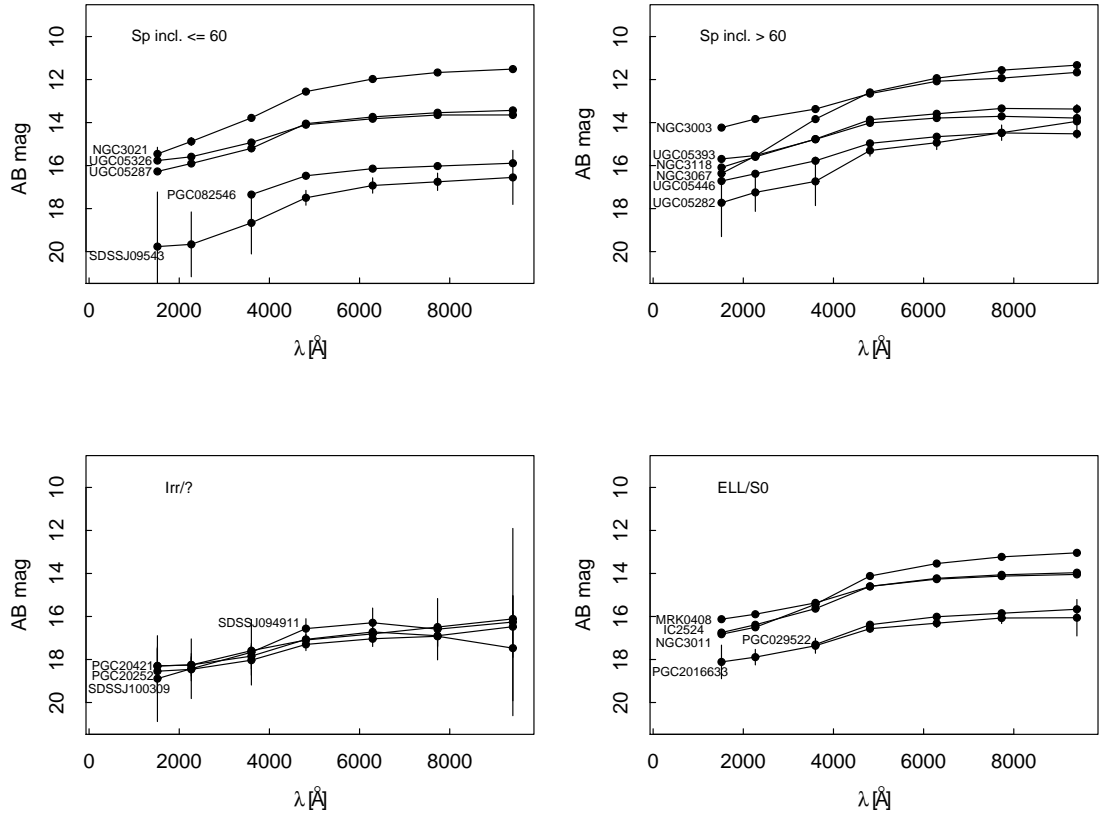}
    \end{center}
           \caption{SEDs (GALEX FUV, NUV, and SDSS {\it u, g, r, i, z} magnitudes) for Spirals with
      inclination $<$ 60$^o$ and $>$ 60$^o$ (left and right top panels), for
      Irregulars and unknown types (left bottom panel) and Ellipticals (right bottom panel) of U268.
      Dots are the integrated magnitudes obtained in the GALEX and SDSS bands as given in \citet[][their Table 6]{Marino13}.  
}
       \label{SEDa}
   \end{figure*}

\begin{figure*}
 \begin{center}
 \includegraphics[width=12cm,angle=-90]{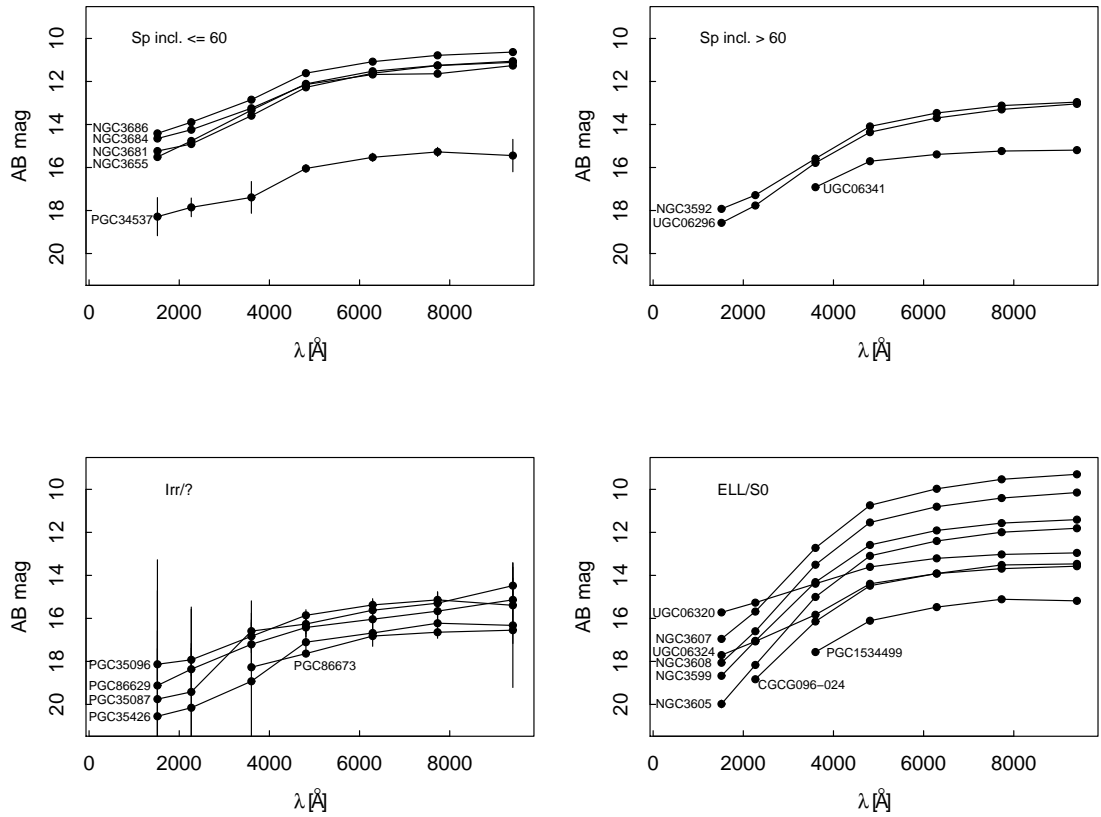}
  \end{center}
      \caption{As in Figure \ref{SEDa} for members of U376.
}
       \label{SEDb}
   \end{figure*}

\section{UV-optical photometry of galaxies in U268 and U376}
   The member galaxies of U268 and  U376  and their main characteristics are listed in Table \ref{tab1}.
Figure \ref{f1} displays color composite  UV and optical images  of 6 members of U376 for which we
 compare results from SED-fitting and from SPH simulations with chemo-photometric implementation. 
 We obtained broad band integrated  magnitudes (AB system) in FUV (1344 -1786 \AA) and near-UV (NUV, 1771 - 2831 \AA)  
from GALEX and in {\it u, g, r, i}, and {\it z}  from Sloan Digital Sky Survey (SDSS) data   
as described in \citet{Marino13}.  
Figures \ref{SEDa} and \ref{SEDb} show the SEDs of the sample galaxies,  arranged by
morphological type. Top panels  show the Spirals with
inclination smaller and larger   than 60$^{\circ}$ respectively, and bottom panels 
Irregulars and Ellipticals/S0s.  
The SEDs of the spiral galaxies are all qualitatively similar  at optical 
wavelengths,  but show a range of slopes
at FUV and NUV  wavelengths, indicative  of a varying relative contribution of the
younger populations.  
Most of the Irregulars also show similar SEDs in optical bands but different slopes in
UV bands. 
 In addition to the recent star formation,
a major factor affecting the  UV fluxes is the extinction by interstellar dust. 
As expected, the SED of the Spirals seen edge-on (Figures \ref{SEDa} and \ref{SEDb},
 top right panels) appears in general 
`redder' than that of  Spirals with low ($<$60$^{\circ}$) inclination
(Figures \ref{SEDa} and \ref{SEDb}, top left panels). 
Including only galaxies with photometric errors $<$0.3 mag in all bands (FUV, NUV, {\it u, g, r, i},
our sample includes  11 and 13 galaxies in U268 and U376, respectively. 
 
\section{Stellar mass and age of the sample galaxies}
  
\subsection{Model grids}

In order to derive the  age and  mass of each galaxy  we computed grids of models  
using the GRASIL  code \citep{Silva98}, for a large range of ages and three different  SFHs.
Extinction must be accounted for, in order to derive correct galaxy parameters.
Internal extinction is included in the GRASIL models; in the SED-fitting process, we also solved for the amount
of possible foreground extinction.
We computed broad-band model magnitudes in the GALEX and SDSS bands, after progressively reddening 
the models by increasing amounts of dust.
We adopted a Salpeter IMF
and the typical parameters regulating the star-formation rate  for spiral, 
irregular and elliptical types respectively,
as given by  \citet{Silva98}\footnote{available  
at  http://adlibitum.oat.ts.astro.it/silva/grasil/modlib/modlib.html.},
and  \citet[][their Table 5]{Marino10}.  
 An important feature of GRASIL
is that the code includes the effect of age-dependent extinction with young stars being 
 more affected by dust. In particular, it takes into account several environments with different dust 
 properties and distributions, such as the AGB envelopes, the diffuse interstellar medium and the
 molecular clouds (MCs).
The geometry of the spiral galaxies in GRASIL is described as a 
superposition of an exponential disc  
and a bulge component with a King profile for Spirals. For Ellipticals and Irregulars 
a spherically symmetric distribution for both stars and gas/dust with a King profile is adopted.

\subsection{Analysis of the observed SEDs with model magnitudes}
From the model spectra we computed synthetic broad-band 
magnitudes in the {\it GALEX}  FUV, NUV  and SDSS {\it u, g, r, i, z}  filters.   
The resulting SEDs at some representative ages have been shown in \citet{Marino10}; their 
  Figure 10    illustrates the effects 
of evolution (left panels) and extinction (right panels).

We used  the model-magnitude grid from a few Myr to 13 Gyr to estimate, by SED fitting  
($\chi^2$ minimization), the best-fit age of the composite population
for each galaxy. The best-fit model scaled to the observed fluxes provides an estimate of the 
current stellar mass, assuming the distances given in \citet{Marino13}, and accounting for extinction. 

\begin{figure*}
 \begin{center}
 \includegraphics[width=6cm]{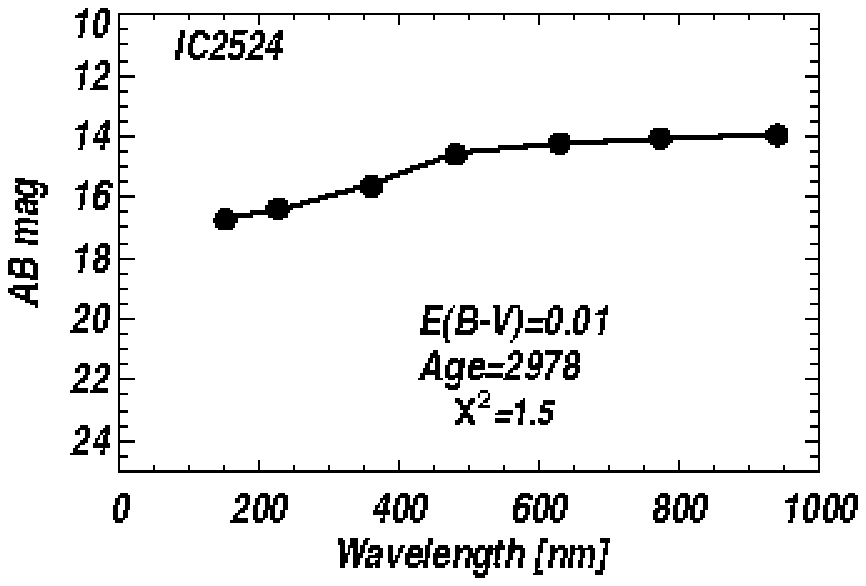} \includegraphics[width=6cm]{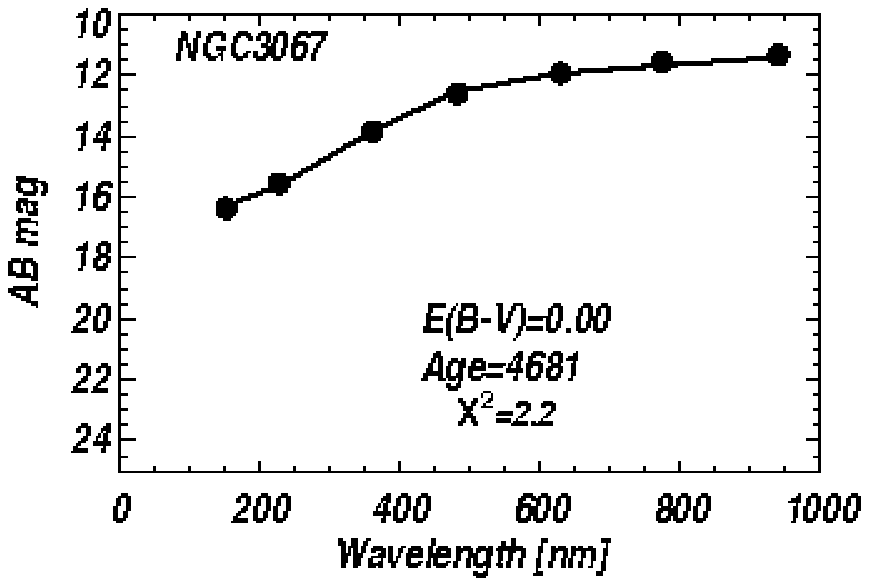}
 \includegraphics[width=6cm]{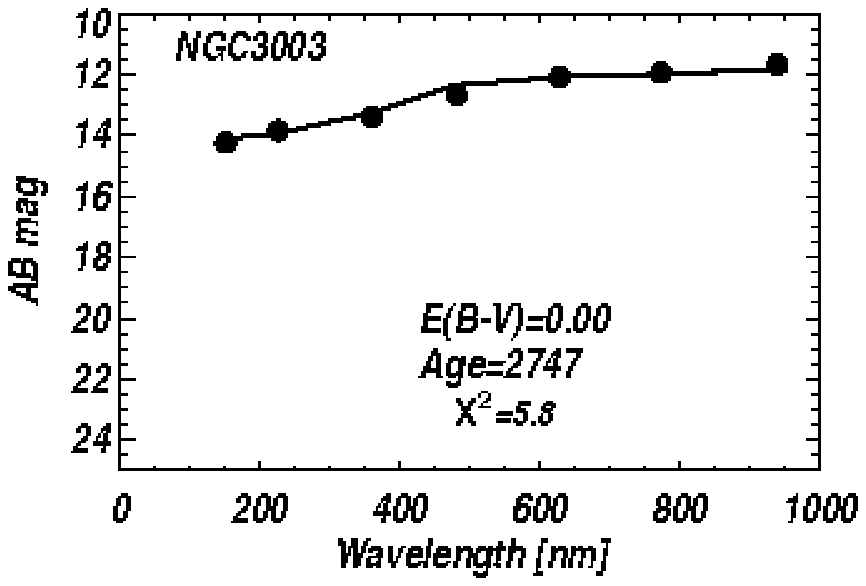}\includegraphics[width=6cm]{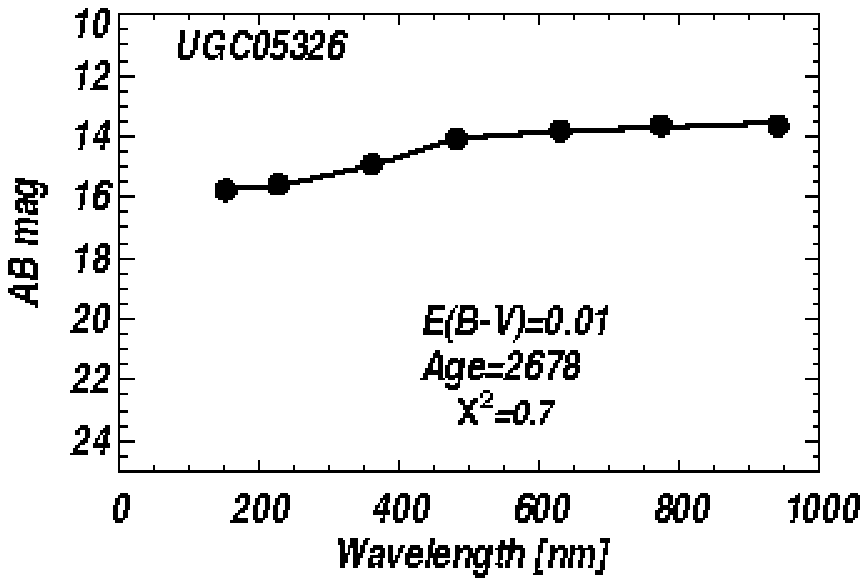} 
         \end{center}
      \caption{FUV, NUV, {\it u, g, r, i, z} SEDs (dots) of IC2524 and NGC 3067 (top panels), 
      NGC 3003 and UGC 5326 (bottom panels)  in U268 with the
       best-fit models (lines).   
        Ages (in Myr) are derived from SED-fitting.  }
       \label{fita}
   \end{figure*}

\begin{figure*}
 \begin{center}
\includegraphics[width=6cm]{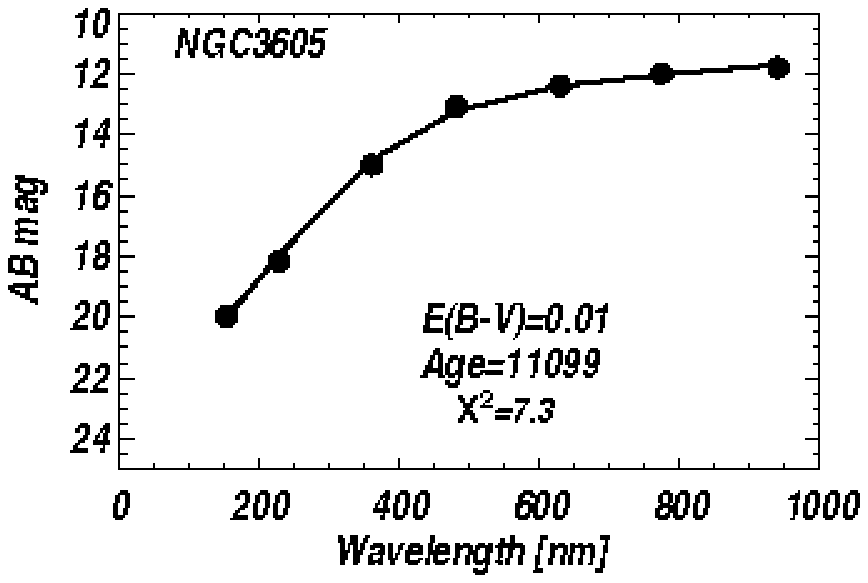} \includegraphics[width=6cm]{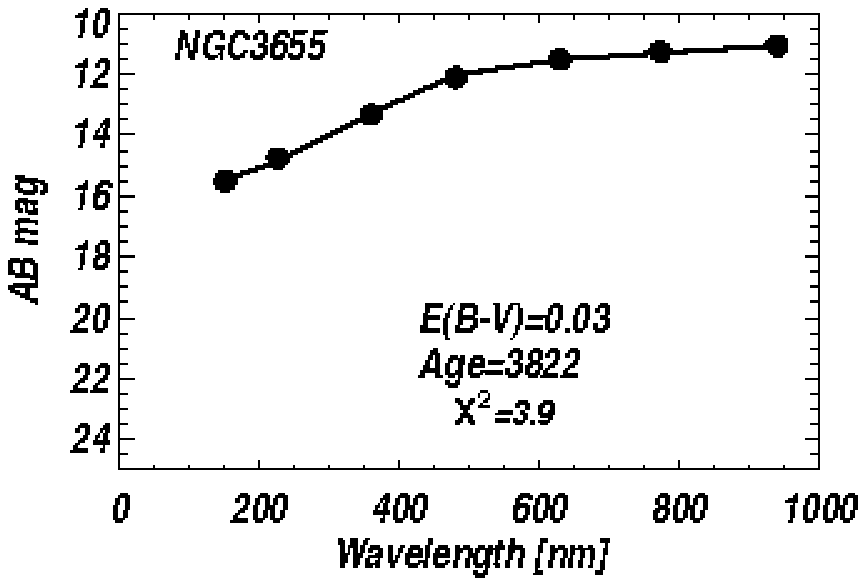}
\includegraphics[width=6cm]{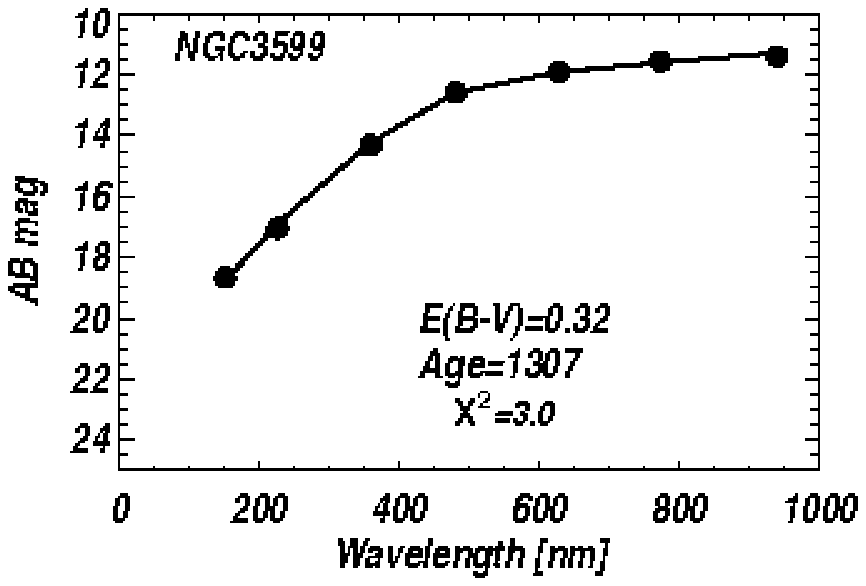} \includegraphics[width=6cm]{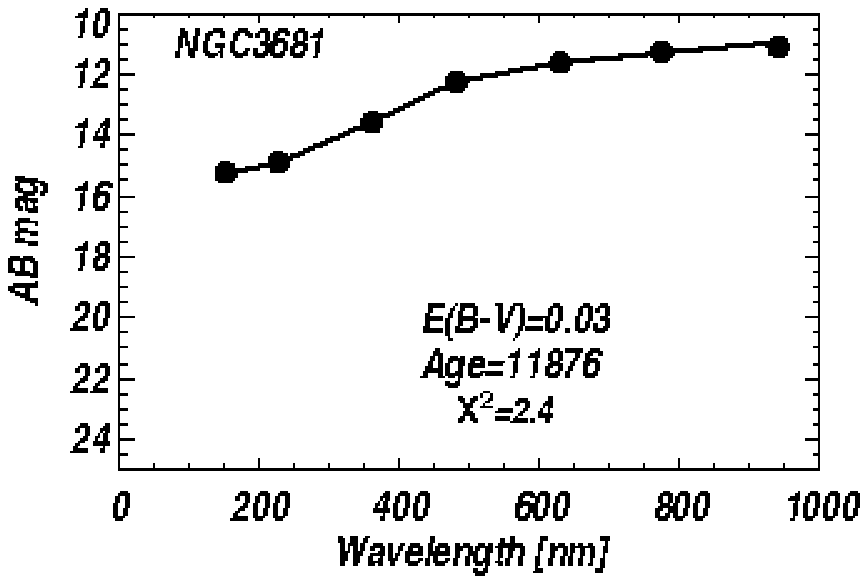}  
  \end{center}
      \caption{As in Figure \ref{fitb} for U376.
}
       \label{fitb}
   \end{figure*}

We performed  SED  fitting in two ways:
first,  assuming foreground extinction as given in Table 1, which is minimal, 
in addition to the internal extinction  estimated by GRASIL  and, second, 
by treating both the age of the population and an additional
extinction component as free parameters.  We compared the two results, and chose the one with
the  minimum $\chi^2$. For all 11 galaxies in U268 and for all but 3 galaxies in U376,
 the best-fit solution confirms the low foreground extinction values ($<$ 0.02 mag) found in previous compilations (Table 1). 
For most Ellipticals/S0s and Spirals  
the models with the  SFH assumed for their respective type   \citep[see Table 5 in][]{Marino10}  give good fits.
The SEDs of three Spirals in U376 (NGC 3681, NGC 3684, and NGC 3686)  and two in U268 (UGC 5393, NGC 3021), 
are also well fitted with the SFH assumed in our grid for
Irregulars. 
 We note that the 3 Spirals in U376  are located outside of the virial radius of the
group (see Figure 6, left panels) and belong to a substructure \citep[][their Fig. 1]{Marino13}.  
U268 does not contain Ellipticals, and the observed SEDs of S0s in this group are all well
reproduced adopting a SFH for Spirals. 
 Figures \ref{fita} and \ref{fitb}  show SED-fitting results for some galaxies with different morphological type belonging 
  to U268 and U376.  
   The age and stellar mass distributions of 11 and 13 galaxies in U268 and  U376 respectively,
for which good SED best-fits has been obtained, are shown in Figure \ref{dist} (right panels).

 We have compared the results obtained with integrated stellar-population models,
with those derived by Mazzei et al. (2013, this book, and in preparation)
using  SPH simulations with chemo-photometric implementation.
Table \ref{sph} gives ages, stellar masses and E(B-V) of a subsample of galaxies  in U376  derived with
the two methods. 
 Stellar masses agree within a factor of $\lesssim$2 in all but 1 case. The ages show large discrepancies
 for 2 galaxies, with SPH  results giving systematically older ages, when discrepant, and
 consequently larger masses. The SFHs derived from the SPH analysis are in agreement with the
  SFHs from SED fitting, except for a few cases, and these explain some discrepant results in age and mass.

\begin{table}
\caption{Stellar mass and age of 7 galaxies in U376 as derived in this work, and in Mazzei et al. (2013). }
 \scriptsize
\begin{tabular}{llllllll}
\hline\hline
\multicolumn{1}{l}{Name} & \multicolumn{1}{l}{Morph.}  &  \multicolumn{1}{l}{Mass$^a$}
 & \multicolumn{1}{l}{Mass$_{SPH}^b$}   & \multicolumn{1}{l}{Age$^a$}   & \multicolumn{1}{l}{Age$_{SPH}^c$}&E(B-V)&E(B-V)$_{SPH}$ \\
 & \multicolumn{1}{l}{Type} & & & & & \\
	& & [10$^{10}\,M_\odot$] & 10$^{10}[M_\odot$]& Gyr  &Gyr&[mag] &[mag]\\
\hline
 NGC~3599&  S0   &1.1(0.6-2.0)  &1.4 - 2.0       &13.0$\pm$0.5    &12.3& 0.01   &0\\
 NGC~3605&  E    &0.6(0.4-1.6)  &1.1 - 1.6      &11.1$\pm$0.5& 12.6&  0.01 &0.01\\
 NGC~3607&  S0   &6.0(3.3-6.8)     & 9.1 -13.0     &13.0$\pm$0.5  & 14&  0.02&    0.04\\
 NGC~3608&  E    &3.0(1.5-6.8)     &9.8 - 14.0    &13.0$\pm$0.5    &13.4& 0.03  &0.01\\
 UCG~06324& S0   &0.2(0.02-0.2)  & 0.2 - 0.3       &3.4$\pm$1.1  & 11 &  0.02   &0\\
 NGC~3655&  Sp   &2.1(0.2-2.2)  &3.5 -  5.0      &3.8$\pm$1.3 & 11.6&0.03  &0.24\\
 NGC~3686&  Sp   &1.2(1.1-2.8)  & 1.8 - 2.6      &7.5$\pm$1.5   &11.6 &0.01 &0.10\\
 \hline
\end{tabular}
$^a$ Values from SED-fitting.  The errors on the ages and masses include not only the uncertainties from $\chi^2$ contours, but the entire range of ages found when different SFHs (for Sp and Irr) are used.  \\  $^b$ Range of mass: within the effective radius and total, from Mazzei et al (2013, in preparation).\\
$^c$ Uncertainties are $\approx$ 0.02 Gyrs.
\label{sph} 
\end{table}

 \section{Kinematic and dynamical properties of U268 and U376.}
 
The virial theorem provides the standard method to estimate the mass of a self-gravitating 
system from dynamical parameters, i.e. positions and velocities of the group members. 
 It is applicable if the system analyzed is in dynamical equilibrium and galaxies trace
 the mass of the system.  

We  derived the kinematic and dynamical properties of the two groups, following
the approach described in \citet{Firth06} and already used in  \citet{Marino10} for
LGG 225.   
This approach allows us to obtain in an homogeneous way the properties of the two groups and
to compare them with those of nearby groups.

\begin{figure*}
\begin{center}
\includegraphics[width=6.5cm,angle=-90]{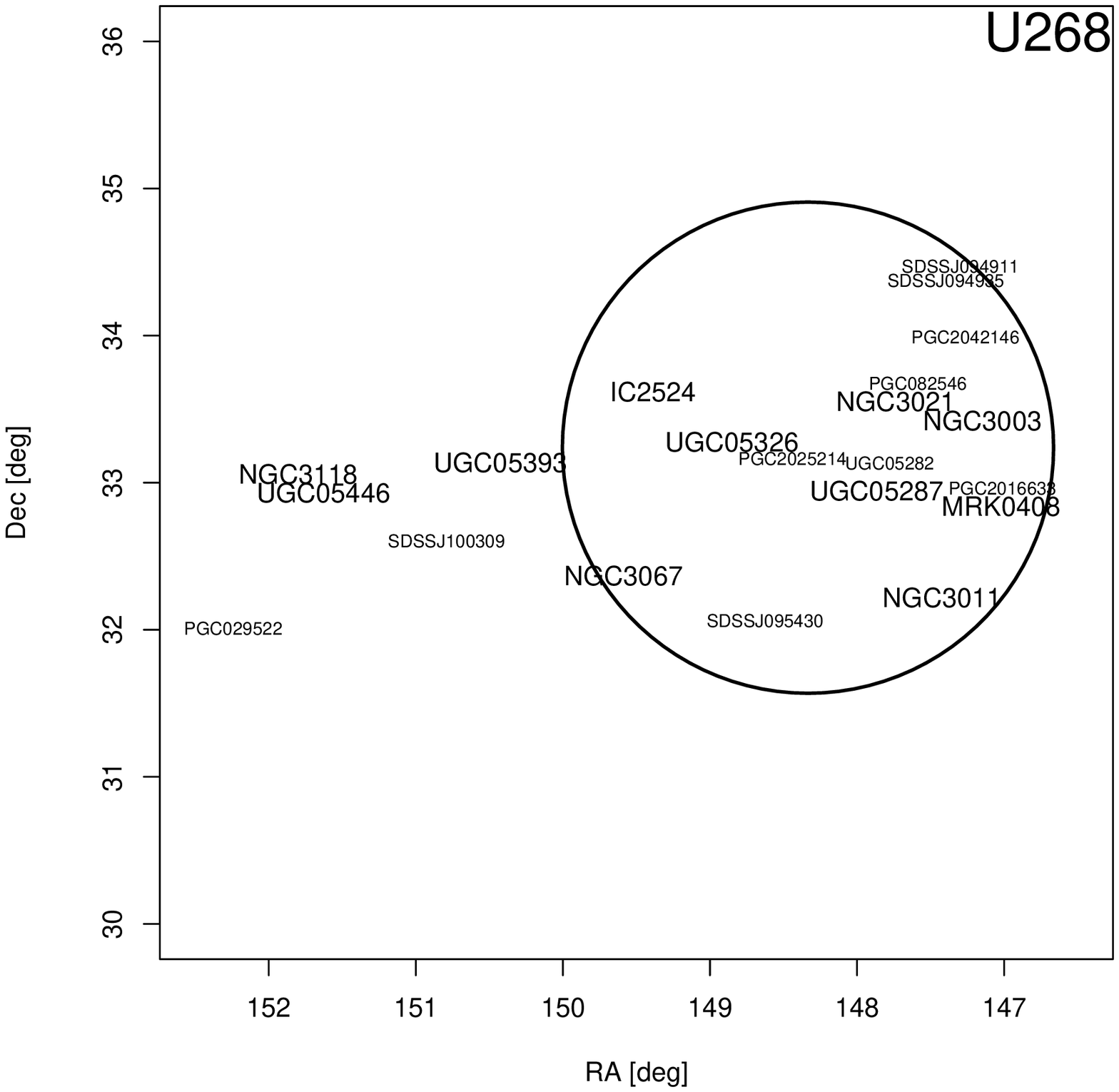}   \includegraphics[width=6.5cm,angle=-90]{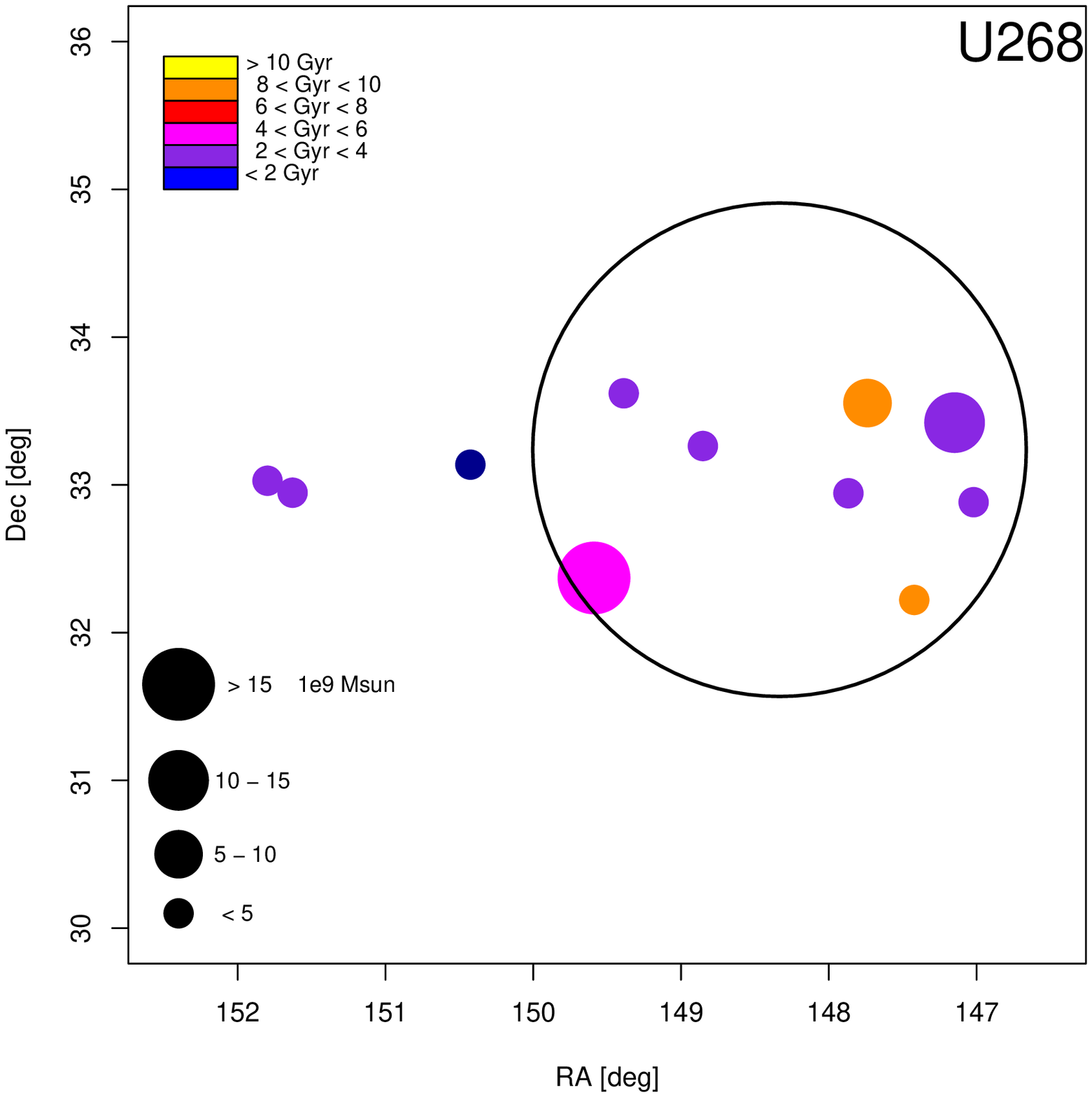}  
\includegraphics[width=6.5cm,angle=-90]{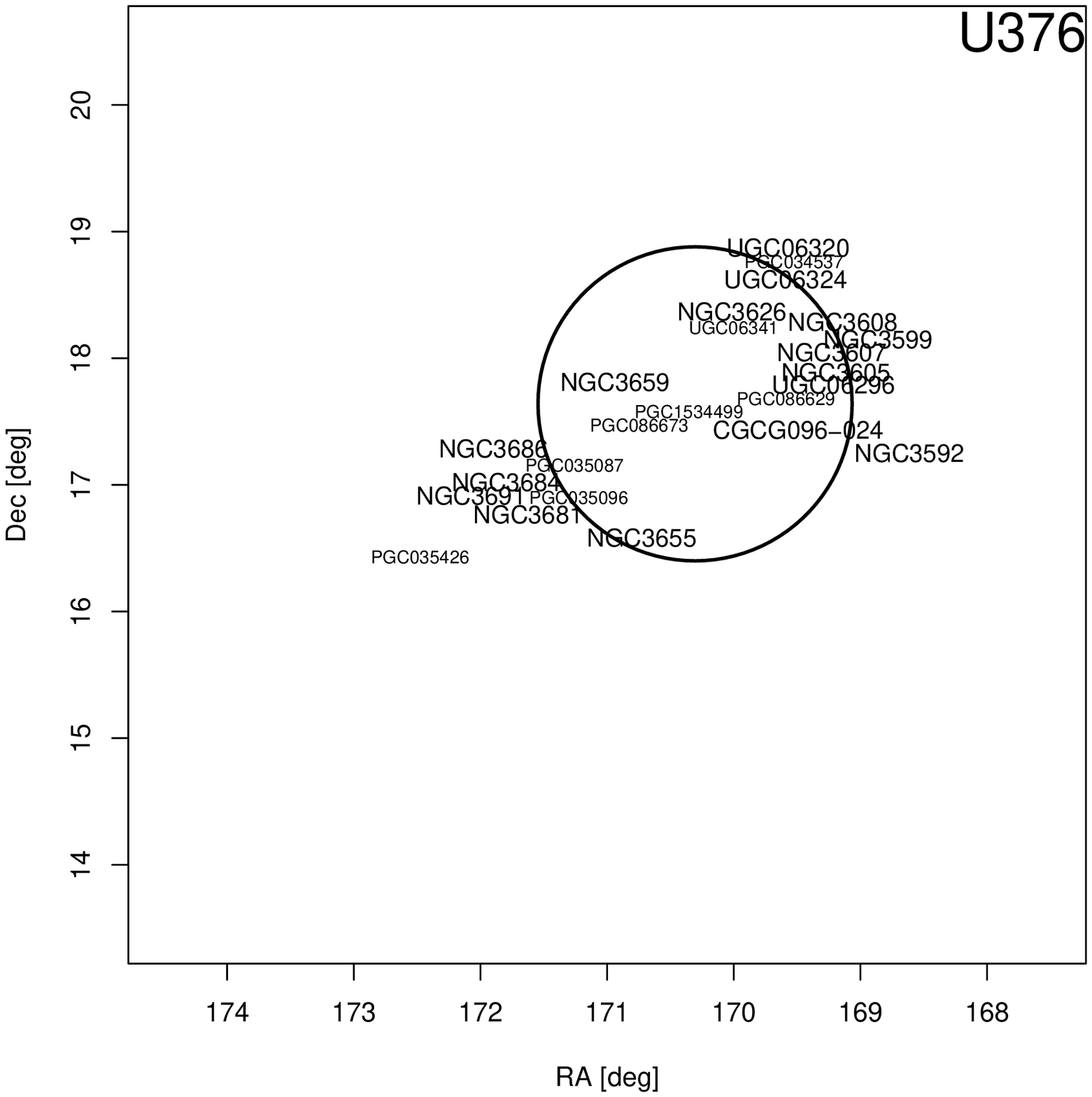}  \includegraphics[width=6.5cm,angle=-90]{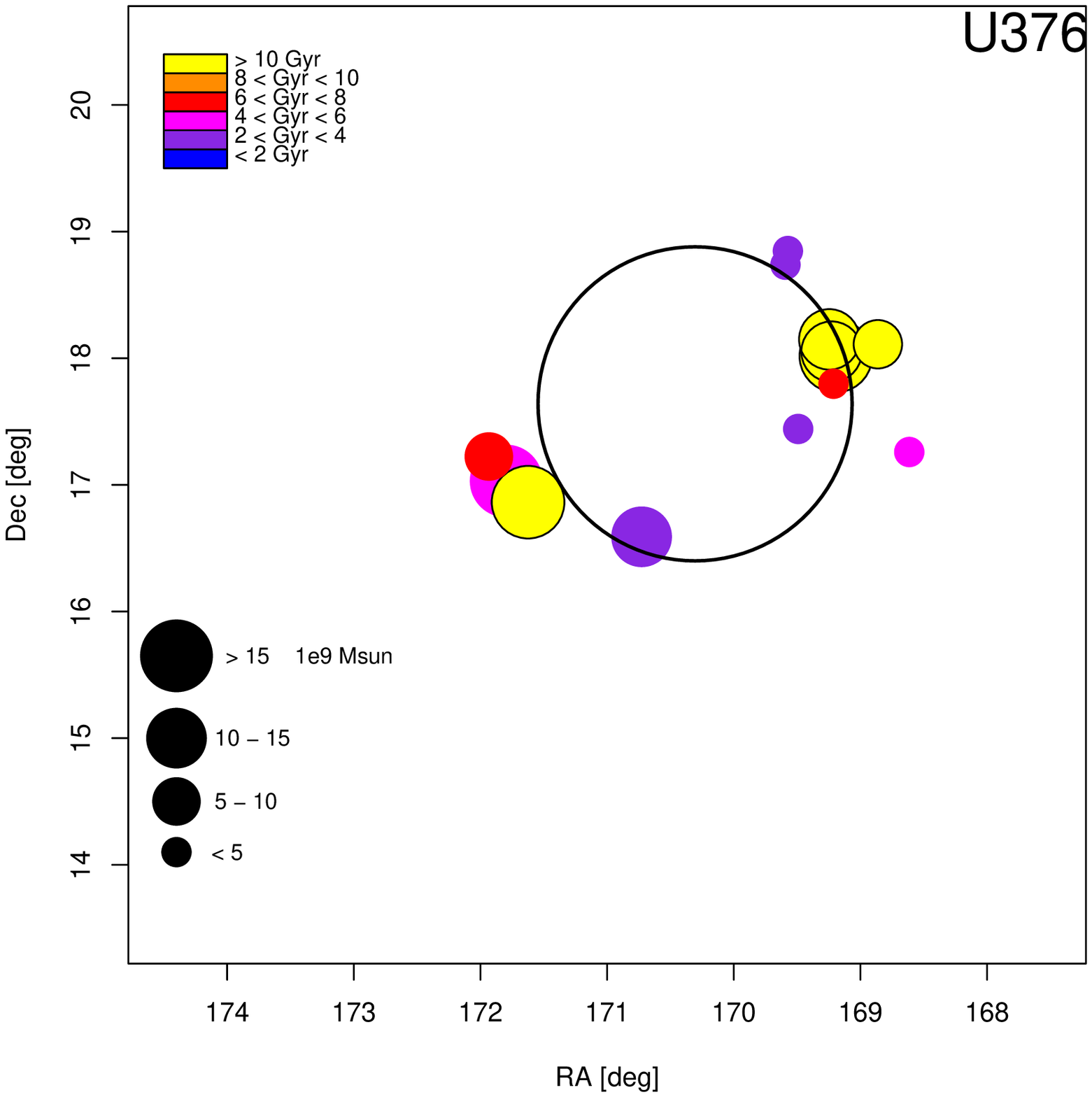} 
\vspace{-0.7cm}
 \end{center}
 \caption{Left: position of  the galaxies in each group, in a square of approximately of 2 $\times$ 2 Mpc$^2$ at their respective distances. 
The circle shows the virial radius around the centre of mass of the group. Right: Age (in Gyr) and stellar mass (in 10$^9$ M$_{\odot}$) distributions of the galaxies in both groups for which reliable SED fitting results has been obtained. } 
	 \label{dist}
\end{figure*}

According to \citet{Firth06}, unaccounted dwarf galaxies do not significantly alter the group velocity dispersions, 
virial mass estimates or crossing times.  The dynamical calculations are based on 
the formulae given in  \citet[][their Table 6]{Firth06}.  The results are summarized in Table~\ref{Dyn}.
Errors in  Table~\ref{Dyn}  have been computed via jackknife simulations \citep[e.g.][]{Efron82}.
The mean velocity of the two groups differs slightly while the velocity dispersion of U376 is  more than three times higher
than that of U268.
 In order to obtain a  measure of the compactness of the three groups, 
we have computed the harmonic mean radius  using
the projected separations r$_{ij}$ between the i-th and j-th group member.
Figure \ref{dist} shows the relative positions of the groups members, with each group scaled to a common distance.
Each square  in Figure \ref{dist} is approximately 2 $\times$ 2 Mpc$^2$. The virial radius  
of U268 (circle in Figure \ref{dist}) is significantly larger than  that of U376, reflecting the more sparse projected
spatial distribution of its members.
   
\begin{table*}
 \caption{Kinematical and dynamical properties of U268 and U376.}
\label{Dyn}
\scriptsize
 \begin{tabular}{lcllllllllllll}
\hline\hline\noalign{\smallskip}
Group & Center  &  V$_{group}$  & Velocity  & Dist.  & Harmonic &Virial  & Projected & Crossing \\ 
 name & of mass &               & dispersion & & radius & mass &  mass &time$\times$H$_0$  \\ 
  & RA ~ Dec &&&&&&\\ 
   & [deg] ~ [deg] &[km/s] &[km/s]& [Mpc] & [Mpc] & [10$^{13}$ M$_{\odot}$] &  10$^{13}$ M$_{\odot}$]   & \\ \hline\hline
\vspace{0.2cm}
 U268       & 148.7295  33.1294   & 1486$^{+9}_{-5}$    &  63$^{+12}_{-2}$   &
 19.8$^{+0.1}_{-0.1}$ & 0.31$^{+0.08}_{-0.02}$ &   0.14$^{+0.05}_{-0.04}$ &  0.59$^{+0.28}_{-0.04}$ &
 0.77$^{+0.09}_{-0.06}$     \\

U376        &170.3058   17.6398   & 1134$^{+21}_{-19}$  &  
225$^{+19}_{-5}$ &   15.1$^{+0.3}_{-0.3}$ &0.21$^{+0.03}_{-0.01}$ & 
1.19$^{+0.20}_{-0.14}$  & 2.98$^{+0.42}_{-0.22}$ &  0.13$^{+0.02}_{-0.01}$    \\
\hline
\end{tabular}
 \end{table*}

The differences between  the virial and projected mass for both groups are quite high (a factor between 3 and 4), 
with the  projected masses higher than the virial masses. This difference
is expected in systems where individual galaxies are close in projection.
Using N-body simulations, \citet{Perea90a} showed that the virial mass estimate is  better than the
projected mass estimate since it is less sensitive to anisotropies  or subclustering. However, it may be affected  
by the presence of interlopers, i.e. unbound galaxies, and by a mass spectrum.
Both factors would cause an overestimation of the group mass, therefore the estimated masses are  
upper limits. 

 Another caveat concerning the virial mass is that the groups may not be virialized \citep[e.g.][]{Ferguson90}.
 The crossing time  is usually compared to the Hubble time to 
determine whether the groups are gravitationally bound systems. The derived crossing time
of U268   
suggests that it  could be unvirialized, while U376  
 could be dynamically relaxed.
The very small velocity dispersion of U268 still suggests that it could be in a pre-virialized phase. 

\section{Summary and conclusions}
We have analyzed the  integrated photometry  of galaxies in  U268 and U376  in the Leo cloud 
with a grid of stellar population synthesis models.   
 
The overall results allow us to characterize the galaxy populations of the two groups:
the composite populations of Spirals have evolutionary times
between 1 and 9  Gyrs for U268 and from 2 to 6 Gyrs for U376, similar to those obtained 
for LGG 225 by \citet{Marino10}.    
The total stellar  mass of the group (sum of the stellar mass estimated for each galaxy)
is $\approx$ 6 $\times 10^{10}$ M$_{\odot}$ for U268 and   $\approx$ 2 $\times 10^{11}$ M$_{\odot}$ for U376.
The narrower age range  of the Spirals  in U376 may be
connected  to their position in the group,  mostly  outside
the virial radius. 
The larger virial radius of U268 includes most of its Spirals and S0s.
The best SED-fits  of these galaxies have been obtained using  SFH typical of Spirals.
 The higher stellar mass and the smaller virial radius  of U376     
 suggest that it is   more evolved than U268,
with LGG 225 in between,   as also suggested  by their 
  Color Magnitude Diagrams \citep[see][their figure 8]{Marino13}.

The stellar masses derived with our SED model-fitting are consistent within a factor of two 
with results from  SPH simulations with chemo-photometric implementation (available for 7 galaxies). 
Ages derived with the two approaches are consistent for 4 galaxies and discrepant for 3,
the  SPH analysis giving older ages for these.  In some cases, more than 
one SED-fitting solution exist (several $\chi^2$ minima), that will be discussed in more detail elsewhere. 
At face value, differences in the discrepant  cases suggest that the SFH of 
galaxies corresponds to a morphological type later than  that adopted
in our model grids. This is not surprising for galaxies located in the group environment in which several studies 
indicate that the galaxy life is quite eventful  \citep[e.g.][]{Annibali07, Marino11a, Marino11b, Marino11c, Rampazzo11}. 
   
 \section*{Acknowledgments}

RR, and AM  acknowledge  partial financial support by contract ASI-INAF  I/009/10/0.


\bibliographystyle{elsarticle-harv.bst}
 \bibliography{amarino}

\end{document}